\documentclass[a4paper,twoside]{article}

\usepackage{epsfig}
\usepackage{subfigure}
\usepackage{calc}
\usepackage{amssymb}
\usepackage{amstext}
\usepackage{amsmath}
\usepackage{amsthm}
\usepackage{multicol}
\usepackage{pslatex}
\usepackage{apalike}
\usepackage{SCITEPRESS}
\usepackage[small]{caption}

\newcommand{\be}{\begin{equation}}
\newcommand{\ee}{\end{equation}}
\begin{document}

\title{A walk in the statistical mechanical formulation of neural networks \subtitle{Alternative routes to Hebb prescription.}}

\author{\authorname{Elena Agliari\sup{1}, Adriano Barra\sup{1}, Andrea Galluzzi\sup{2}, Daniele Tantari\sup{2} and Flavia Tavani\sup{3}}
\affiliation{\sup{1}Dipartimento di Fisica, Sapienza Universita' di Roma.}
\affiliation{\sup{2}Dipartimento di Matematica, Sapienza Universita' di Roma.}
\affiliation{\sup{3}Dipartimento SBAI (Ingegneria), Sapienza Universita' di Roma.}
\email{\{elena.agliari, adriano.barra\}@roma1.infn.it, \{galluzzi, tantari\}@mat.uniroma1.it, flavia.tavani@sbai.uniroma1.it}
}


\abstract{Neural networks are nowadays both powerful operational tools (e.g., for pattern recognition, data mining, error correction codes) and complex theoretical models on the focus of scientific investigation. As for the research branch, neural networks are handled and studied by psychologists, neurobiologists, engineers, mathematicians and theoretical physicists. In particular, in theoretical physics, the key instrument for the quantitative analysis of neural networks is statistical mechanics.
From this perspective, here, we first review attractor networks: starting from ferromagnets and spin-glass models, we discuss the underlying philosophy and we recover the strand paved by Hopfield, Amit-Gutfreund-Sompolinky.
One step forward, we highlight the structural equivalence between Hopfield networks (modeling retrieval) and Boltzmann machines (modeling learning), hence realizing a deep bridge linking two inseparable aspects of biological and robotic spontaneous cognition.
As a sideline, in this walk we derive two alternative (with respect to the original Hebb proposal) ways to recover the Hebbian paradigm, stemming from ferromagnets and from spin-glasses, respectively. Further, as these notes are thought of for an Engineering audience, we highlight also the mappings between ferromagnets and operational amplifiers  and between antiferromagnets and  flip-flops (as neural networks -built by op-amp and flip-flops- are particular spin-glasses and the latter are indeed combinations of ferromagnets and antiferromagnets), hoping that such a bridge plays as a concrete prescription to capture the beauty of robotics from the statistical mechanical perspective.}

\onecolumn \maketitle \normalsize \vfill

\section{Introduction}

Neural networks are such a fascinating field of science that its development is the result of contributions and efforts from an incredibly large variety of scientists, ranging from \emph{engineers} (mainly involved in electronics and robotics) \cite{1,2}, \emph{physicists} (mainly involved in statistical mechanics and stochastic processes) \cite{3,4}, and \emph{mathematicians} (mainly working in logics and graph theory) \cite{5,6} to \emph{(neuro) biologists} \cite{7,8} and \emph{(cognitive) psychologists} \cite{9,10}.

Tracing the genesis and evolution of neural networks is very difficult, probably due to the broad meaning they have acquired along the years\footnote{Seminal ideas regarding automation are already in the works of Lee during the XIIX century, if not even back to Descartes, while more modern ideas regarding {\em spontaneous cognition}, can be attributed to A. Turing \cite{turing} and J. Von Neumann \cite{neumann} or to the join efforts of M. Minsky and S. Papert \cite{papert}, just to cite a few.}; scientists closer to the robotics branch often refer to the W. McCulloch and W. Pitts model of perceptron \cite{11} \footnote{Note that the first ''transistor'', crucial to switch from analogical to digital processing, was developed only in 1948 \cite{12}.}, or the F. Rosenblatt version \cite{13}, while researchers closer to the neurobiology branch adopt D. Hebb's work as a starting point \cite{14}. On the other hand, scientists involved in statistical mechanics, that joined the community in relatively recent times, usually refer to the seminal paper by Hopfield \cite{15} or to the celebrated work by Amit Gutfreund Sompolinky \cite{3}, where the statistical mechanics analysis of the Hopfield model is effectively carried out.

Whatever the reference framework, at least 30 years elapsed since neural networks entered in the theoretical physics research and much of the former results can now be re-obtained or re-framed in modern approaches and made much closer to the engineering counterpart, as we want to highlight in the present work. In particular, we show that toy models for paramagnetic-ferromagnetic transition \cite{16} are natural prototypes for the autonomous storage/retrieval of information patterns and play as operational amplifiers in electronics. Then, we move further analyzing the capabilities of glassy systems (ensembles of ferromagnets and antiferromagnets) in storing/retrieving extensive numbers of patterns so to recover the Hebb rule for learning \cite{14} in two different ways (the former guided by ferromagnetic intuition, the latter guided by glassy counterpart), both far from the original route contained in his milestone {\em The Organization of Behavior}. Finally, we will give prescription to map these glassy systems in ensembles of amplifiers and inverters (thus flip-flops) of the engineering counterpart so to offer a concrete bridge between the two communities of theoretical physicists working with complex systems and engineers working with robotics and information processing.

As these notes are intended for non-theoretical-physicists, we believe that they can constitute a novel perspective on a by-now classical theme and that they could hopefully excite curiosity toward statistical mechanics in nearest neighbors scientists like engineers whom these proceedings are addressed to.

\section{Statistical mechanics: Microscopic dynamics obeying entropy maximization}

Hereafter we summarize the fundamental steps that led theoretical physicists towards artificial intelligence; despite this parenthesis may look rather distant from neural network scenarios, it actually allows us to outline and to historically justify the physicists perspective.

Statistical mechanics aroused in the last decades of the XIX century thanks to its founding fathers Ludwig Boltzmann, James Clarke Maxwell and Josiah Willard Gibbs \cite{17}. Its ``solely'' scope (at that time) was to act as a theoretical ground of the already existing empirical thermodynamics, so to reconcile its noisy and irreversible behavior with a deterministic and time reversal microscopic dynamics.  While trying to get rid of statistical mechanics in just a few words  is almost meaningless, roughly  speaking its functioning may be summarized via toy-examples as follows. Let us consider a very simple system, e.g. a perfect gas: its molecules obey a Newton-like microscopic dynamics (without friction -as we are at the molecular level- thus time-reversal as dissipative terms in differential equations capturing system's evolution are coupled to odd derivatives) and, instead of focusing on each particular trajectory for characterizing the state of the system, we define order parameters (e.g. the density) in terms of microscopic variables (the particles belonging to the gas).   By averaging their evolution over suitably probability measures, and imposing on these averages energy minimization and entropy maximization, it is possible to infer the macroscopic behavior in agreement with thermodynamics, hence bringing together the microscopic deterministic and time reversal mechanics with the macroscopic strong dictates stemmed by the second principle (i.e. arrow of time coded in the entropy growth).
Despite famous attacks to Boltzmann theorem (e.g. by Zermelo or Poincar\'{e}) \cite{18}, statistical mechanics  was immediately recognized as a deep and powerful bridge linking microscopic dynamics of a system's constituents with (emergent) macroscopic properties shown by the system itself, as exemplified by the equation of state for \emph{perfect gases} obtained by considering an Hamiltonian for a single particle accounting for the kinetic contribution only \cite{17}.

One step forward beyond the perfect gas, Van der Waals and Maxwell in their pioneering works focused on \emph{real gases} \cite{19}, where particle interactions were finally considered by introducing a non-zero potential in the microscopic Hamiltonian describing the system. This extension implied fifty-years of deep changes in the theoretical-physics perspective in order to be able to face new classes of questions. The remarkable reward lies in a theory of phase transitions where the focus is no longer on details regarding the system constituents, but rather on the characteristics of their interactions.
Indeed, phase transitions, namely abrupt changes in the macroscopic state of the whole system, are not due to the particular system considered, but are primarily due to the ability of its constituents to perceive interactions over the thermal noise. For instance, when considering a system made of by a large number of water molecules, whatever the level of resolution to describe the single molecule (ranging from classical to quantum), by properly varying the external tunable parameters
(e.g. the temperature\footnote{We chose the temperature here (as an example of tunable parameter) because in neural networks we will deal with white noise affecting the system. Analogously, in condensed matter, disorder is introduced by thermal noise, namely temperature. There is a deep similarity between them. In stochastic processes, prototype for white noise generators are random walkers, whose continuous limits are Gaussians, namely  just the solutions of the Fourier equation for diffusion. However, the same celebrated equation holds for temperature spread too, indeed the latter is related to the amount of exchanged heat by the system under consideration, necessary for entropy's growth \cite{19,20}. Hence we have the first equivalence: white noise in neural networks mirrors thermal noise in structure of matter.}), this {\em system} eventually changes its state from liquid to vapor (or solid, depending on parameter values); of course, the same applies generally to liquids.

The fact that the macroscopic behavior of a system may spontaneously show {\em cooperative, emergent} properties, actually hidden in its microscopic description and not directly deducible when looking at its components alone, was definitely appealing in neuroscience. In fact, in the 70s neuronal dynamics along axons, from dendrites to synapses, was already rather clear (see e.g. the celebrated book by Tuckwell \cite{21}) and not too much intricate than circuits that may arise from basic human creativity: remarkably simpler than expected and certainly trivial with respect to overall cerebral functionalities like learning or computation, thus the aptness of a {\em thermodynamic formulation} of neural interactions -to {\em reveal} possible emergent capabilities- was immediately pointed out, despite the route was not clear yet.

Interestingly, a big step forward to this goal was prompted by problems stemmed from condensed matter. In fact, quickly theoretical physicists realized that the purely kinetic Hamiltonian, introduced for perfect gases (or Hamiltonian with mild potentials allowing for real gases), is no longer suitable for solids, where atoms do not move freely and the main energy contributions are from potentials. An ensemble of harmonic oscillators (mimicking atomic oscillations of the nuclei around their rest positions) was the first scenario for understanding condensed matter: however, as experimentally revealed by crystallography, nuclei are arranged according to regular lattices hence motivating mathematicians in study periodical structures to help physicists in this modeling, but merging statistical mechanics with lattice theories resulted  soon in practically intractable models\footnote{For instance the famous Ising model \cite{22}, dated $1920$ (and curiously invented by Lenz) whose properties are known in dimensions one and two, is still waiting for a solution in three dimensions.}.

As a paradigmatic example, let us consider the one-dimensional Ising model, originally introduced to investigate magnetic properties of matter:  the generic, out of $N$, nucleus labeled as $i$ is schematically represented by a spin $\sigma_i$, which can assume only two values ($\sigma_i=-1$, spin down and $\sigma_i=+1$, spin up); nearest neighbor spins interact reciprocally through positive (i.e. ferromagnetic) interactions $J_{i,i+1}>0$, hence the Hamiltonian of this system can be written as $H_N(\sigma) \propto -\sum_i^N J_{i,i+1}\sigma_i \sigma_{i+1} -h \sum_i^N \sigma_i$, where $h$ tunes the external magnetic field and the minus sign in front of each term of the Hamiltonian ensures that spins try to align with the external field and to get parallel each other in order to fulfill the minimum energy principle.
\newline
Clearly this model can trivially be extended to higher dimensions, however, due to prohibitive difficulties in facing the topological constraint of considering nearest neighbor interactions only, soon shortcuts were properly implemented to turn around this path. It is just due to an effective shortcut, namely the so called ``mean field approximation'', that statistical mechanics approached complex systems and, in particular, artificial intelligence.

\section{The route to complexity: The role of mean field limitations}
As anticipated, the ``mean field approximation'' allows overcoming prohibitive technical difficulties owing to the underlying lattice structure. This consists in extending the sum on nearest neighbor couples (which are $\mathcal{O}(N)$) to include all possible couples in the system (which are $\mathcal{O}(N^2$)), properly rescaling the coupling ($J \to $J/N$)$ in order to keep thermodynamical observables linearly extensive. If we consider a ferromagnet built of by $N$ Ising spins $\sigma_i = \pm 1$ with $i \in (1,...,N)$, we can then write
\be\label{CWH}
H_{N}(\sigma|J) = -\frac1N \sum_{i<j}^{N,N} J_{ij}\sigma_i \sigma_j \sim - \frac{1}{2N}\sum_{i,j}^{N,N}\sigma_i\sigma_j,
\ee
where in the last term we neglected the diagonal term $(i=j)$ as it is irrelevant for large $N$. From a topological perspective the mean-field approximation equals to abandon the lattice structure in favor to a complete graph (see Fig.~$1$).
When the coupling matrix has only positive entries, e.g. $P(J_{ij}) = \delta(J_{ij}-J)$, this model is named Curie-Weiss model and acts as the simplest microscopic Hamiltonian able to describe the paramagnetic-ferromagnetic transitions experienced by materials when temperature is properly lowered. An external (magnetic) field $h$ can be accounted for by adding in the Hamiltonian an extra term $\propto - h \sum_i^N \sigma_i$.

\begin{figure} \label{fig:grafi}
 \includegraphics[width=0.45\textwidth]{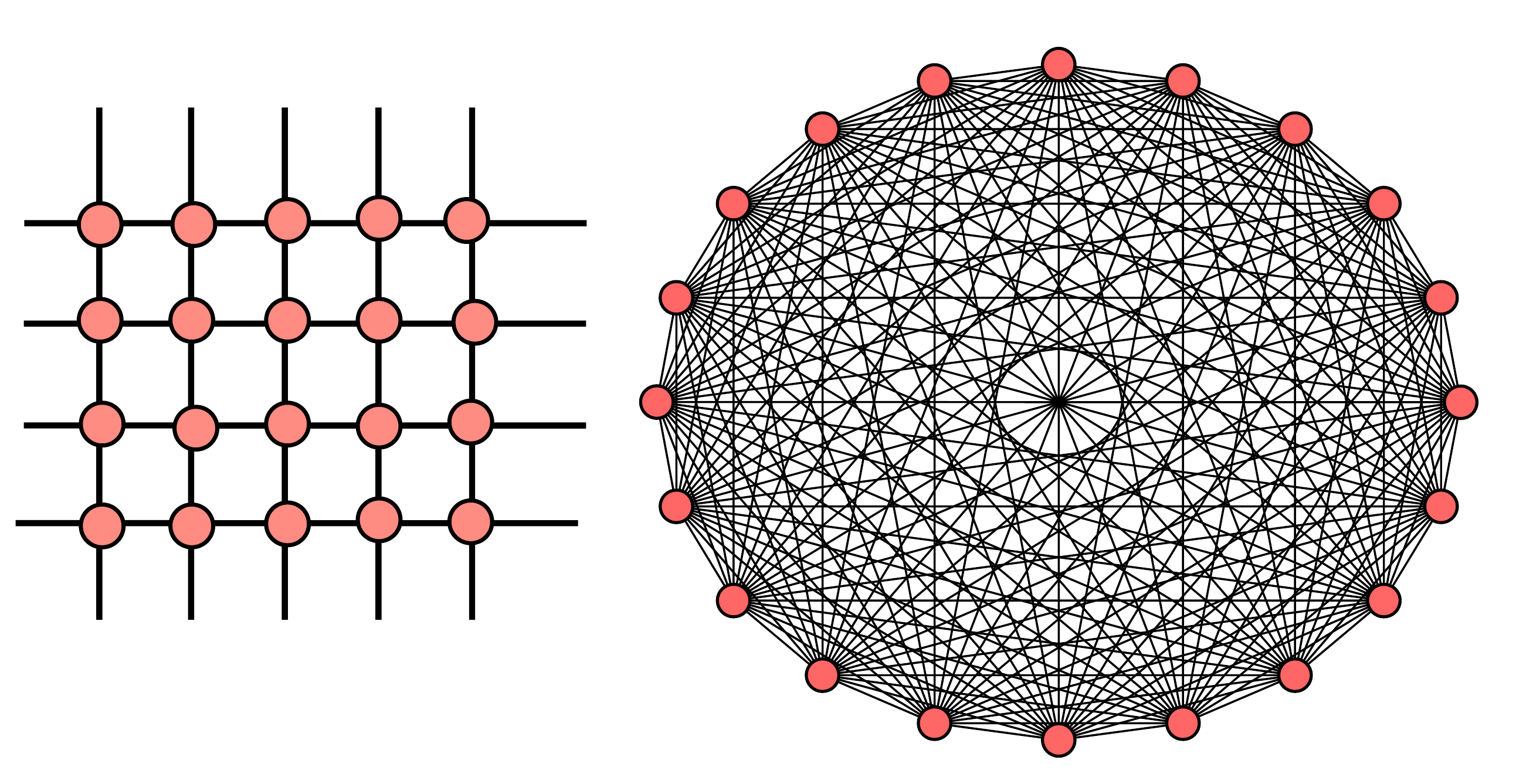}
 \caption{Example of regular lattice (left) and complete graph (right) with $N=20$ nodes. In the former only nearest-neighbors are connected in such a way that the number of links scales linearly with $N$, while in the latter each node is connected with all the remaining $N-1$ in such a way that the number of links scales quadratically with $N$.}
\end{figure}

According to the principle of minimum energy, the two-body interaction appearing in the Hamiltonian in Eq.~\ref{CWH} tends to make spins parallel with each other and aligned with the external field if present. However, in the presence of noise (i.e. if temperature $T$ is strictly positive), maximization of entropy must also be taken into account. When the noise level is much higher than the average energy (roughly, if $T \gg J$), noise and entropy-driven disorder prevail and spins are not able to ``feel'' reciprocally; as a result, they flip randomly and the system behaves as a {\em paramagnet}. Conversely, if noise is not too loud, spins start to interact possibly giving rise to a phase transition; as a result the system globally rearranges its structure orientating all the spins in the same direction, which is the one selected by the external field if present, thus we have a {\em ferromagnet}.

In the early '70 a scission occurred in the statistical mechanics community: on the one side ``pure physicists" saw mean-field approximation as a merely bound to bypass in order to have satisfactory pictures of the structure of matter and they succeeded in working out iterative procedures to embed statistical mechanics in (quasi)-three-dimensional reticula, yielding to the \emph{renormalization group} developed by Kadanoff and Wilson in America \cite{23} and Di-Castro and Jona-Lasinio in Europe \cite{24}; this proliferative branch gave then rise to superconductivity, superfluidity \cite{25} and many-body problems in condensed matter \cite{26}.
\newline
Conversely, from the other side, the mean-field approximation acted as a breach in the wall of complex systems: a thermodynamical investigation of phenomena occurring on general structures lacking Euclidean metrics (e.g. Erdos-Renyi graphs \cite{27,27bis}, small-world graphs \cite{28,28bis}, diluted, weighted graphs \cite{troia}) was then possible.

In general, as long as the summations run over all the indexes (hence mean-field is retained), rather complex coupling patterns can be solved (see e.g., the striking Parisi picture of mean-field glassy systems \cite{MPV}) and this paved the strand to complex system analysis by statistical mechanics, whose investigation largely covers neural networks too.

\section{Serial processing}

Hereafter we discuss how to approach neural networks from models mimicking ferromagnetic transition. In particular, we study the Curie-Weiss model and we show how it can store one pattern of information and then we bridge its input-output relation (called {\em self-consistency}) with the transfer function of an operational amplifier. Then, we notice that such a stored pattern has a very peculiar structure which is hardly {\em natural}, but we will overcome this (fake) flaw by introducing a gauge variant known as Mattis model. This scenario can be looked at as a primordial neural network and we discuss its connection with biological neurons and operational amplifiers. The successive step consists in extending, through elementary thoughts, this picture in order to include and store several patterns. In this way, we recover, via the first alternative route (w.r.t. the original one by Hebb), both the Hebb rule for synaptic plasticity and, as a corollary, the Hopfield model for neural networks too that will be further analyzed in terms of flip-flops and information storage.

\subsection{Storing the first pattern: Curie-Weiss paradigm.}

The statistical mechanical analysis of the Curie-Weiss model (CW) can be summarized as follows: Starting from a microscopic formulation of the system, i.e. $N$ spins labeled as $i,j,...$, their pairwise couplings $J_{ij}\equiv J$, and possibly an external field $h$, we derive an explicit expression for its (macroscopic) free energy $A(\beta)$. The latter is the effective energy, namely the difference between the internal energy $U$, divided by the temperature $T=1/\beta$, and the entropy $S$, namely $A(\beta) = S - \beta U$, in fact, $S$ is the ``penalty'' to be paid to the Second Principle for using $U$ at noise level $\beta$. We can therefore link macroscopic free energy with microscopic dynamics via the fundamental relation
\be \label{alpha}
A(\beta) = \lim_{N \to \infty} \frac1N \ln \sum_{\{ \sigma \}}^{2^N} \exp\left[ -\beta H_N(\sigma|J,h)\right],
\ee
where the sum is performed over the set $\{ \sigma\}$ of all $2^N$ possible spin configurations, each weighted by the Boltzmann factor $\exp[-\beta H_N(\sigma|J,h)]$ that tests the likelihood of the related configuration.
From expression (\ref{alpha}), we can derive the whole thermodynamics and in particular phase-diagrams, that is, we are able to discern regions in the space of tunable parameters (e.g. temperature/noise level) where the system behaves as a paramagnet or as a ferromagnet.
\newline
Thermodynamical averages, denoted with the symbol $\langle . \rangle$, provide for a given observable the expected value, namely the value to be compared with measures in an experiment. For instance, for the magnetization $m(\sigma) \equiv \sum_{i=1}^N \sigma_i /N$ we have
\be
\langle m (\beta) \rangle = \frac{\sum_{\sigma} m(\sigma) e^{-\beta H_N(\sigma|J)}}{\sum_{\sigma}e^{-\beta H_N(\sigma|J)}}.
\ee
When $\beta \to \infty$ the system is noiseless (zero temperature) hence spins feel reciprocally without errors and the system behaves ferromagnetically ($|\langle m\rangle|\to 1$), while when $\beta \to 0$ the system behaves completely random (infinite temperature), thus interactions can not be felt and the system is a paramagnet ($\langle m \rangle\to 0$). In between a phase transition happens.

In the Curie-Weiss model the magnetization works as {\em order parameter}: its thermodynamical average is zero when the system is in a paramagnetic (disordered) state ($\to \langle m \rangle =0$), while it is different from zero in a ferromagnetic state (where it can be either positive or negative, depending on the sign of the external field). Dealing with order parameters allows us to avoid managing an extensive number of variables $\sigma_i$, which is practically impossible and, even more important, it is not strictly necessary.

Now, an explicit expression for the free energy in terms of $\langle m \rangle$ can be obtained carrying out summations in eq. \ref{alpha} and taking the {\em thermodynamic limit} $N \to \infty$ as
\be
A(\beta) = \ln 2 + \ln\cosh [ \beta (J \langle m \rangle + h) ]  - \frac{\beta J}{2}\langle m \rangle^2.
\ee
In order to impose thermodynamical principles, i.e. energy minimization and entropy maximization, we need to find the extrema of this expression with respect to $\langle m \rangle$ requesting $\partial_{\langle m(\beta) \rangle}A(\beta)=0$. The resulting expression is called the \emph{self-consistency} and it reads as
\be\label{self}
\partial_{\langle m \rangle}A(\beta)=0 \Rightarrow \langle m \rangle = \tanh[\beta (J \langle m \rangle + h)].
\ee
This expression returns the average behavior of a spin in a magnetic field. In order to see that a phase transition between paramagnetic and ferromagnetic states actually exists, we can fix $h=0$ (and pose $J=1$ for simplicity) and expand the r.h.s. of eq.~\ref{self} to get
\be \label{crit}
\langle m \rangle \propto \pm \sqrt{\beta J - 1}.
\ee
Thus, while the noise level is higher than one ($\beta < \beta_c \equiv 1$ or $T > T_c \equiv 1)$ the only solution is $\langle m \rangle =0$, while, as far as the noise is lowered below its critical threshold $\beta_c$, two different-from-zero branches of solutions appear for the magnetization and the system becomes a ferromagnet (see Fig.$2$). The branch effectively chosen by the system usually depends on the sign of the external field or boundary fluctuations: $\langle m \rangle >0$ for $h>0$ and  vice versa for $h<0$.

Clearly, the lowest energy minima correspond to the two configurations with all spins aligned, either upwards ($\sigma_i = +1 , \forall i$) or downwards ($\sigma_i = -1 , \forall i$), these configurations being symmetric under spin-flip $\sigma_i \to - \sigma_i$. Therefore, the thermodynamics of the Curie-Weiss model is solved: energy minimization tends to align the spins (as the lowest energy states are the two ordered ones), however entropy maximization tends to randomize the spins (as the higher the entropy, the most disordered the states,  with half spins up and half spins down): the interplay between the two principles is driven by the level of noise introduced in the system and this is in turn ruled by the tunable parameter $\beta \equiv 1/T$ as coded in the definition of free energy.
\begin{figure}
 \includegraphics[width=0.4\textwidth]{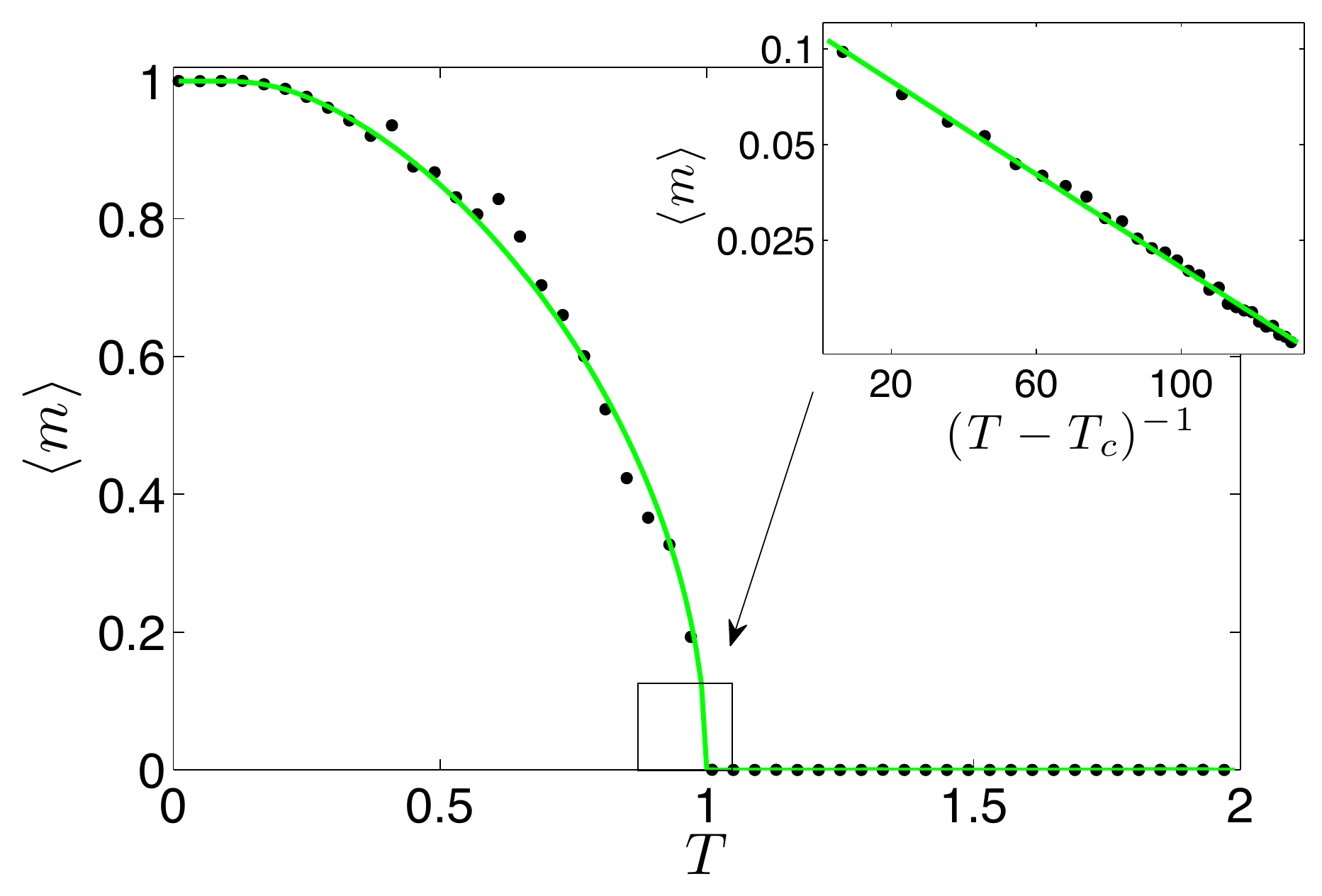}
 \caption{Average magnetization $\langle m \rangle$ versus temperature $T$ for a Curie-Weiss model in the absence of field ($h=0$). The critical temperature $T_c=1$ separates a magnetized region ($|\langle m \rangle| >0$, only one branch shown) from a non-magnetized region ($\langle m \rangle =0$). The box zooms over the critical region (notice the logarithmic scale) and highlights the power-law behavior $m \sim (T-T_c)^{\beta}$, where $\beta=1/2$ is also referred to as critical exponent (see also eq.~\ref{crit}). Data shown here ($\bullet$) are obtained via Monte Carlo simulations for a system of $N=10^5$ spins and compared with the theoretical curve (solid line).}
\end{figure}

A crucial bridge between condensed matter and neural network could now be sighted: One could think at each spin as a basic neuron, retaining only its ability to spike such that $\sigma_i=+1$ and $\sigma_i=-1$ represent firing and quiescence, respectively, and associate to each equilibrium configuration of this spin system a {\em stored pattern} of information. The reward is that, in this way, the spontaneous (i.e. thermodynamical) tendency of the network to relax on free-energy minima can be related to the spontaneous retrieval of the stored pattern, such that the cognitive capability emerges as a natural consequence of physical principles: we well deepen this point along the whole paper.

\subsection{The Curie-Weiss model and the (saturable) operational amplifier}

Let us now tackle the problem by another perspective and  highlight a structural/mathematical similarity in the world of electronics:  the plan is to compare self-consistencies in statistical mechanics and transfer functions in electronics so to reach a unified description for these systems. Before proceeding, we recall a few basic concepts. The operational amplifier, namely a solid-state integrated circuit (transistor), uses feed-back regulation to set its functions as sketched in Fig.~$3$ (insets): there are two signal inputs (one positive received (+) and one inverted, thus
negative received (-)), two voltage supplies ($V_{sat}$, -$V_{sat}$) -where ``sat'' stands for {\em saturable} - and an output ($V_{out}$). An ideal amplifier is the {\em linear} approximation of the saturable one (technically the voltage at the input collectors is thought constant so that no current flows inside the transistor, and Kirchoff rules apply straightforwardly): keeping the symbols of Fig.~$3$, where $R_{in}$ stands for the input resistance while $R_f$ represents the feed-back resistance, $i_+=i_- =0$ and assuming $R_{in} = 1\Omega$ -without loss of generality as only the ratio $R_{f}/R_{in}$ matters- the following transfer function is achieved:
\be
V_{out}=G V_{in} = (1 + R_{f}) V_{in},
\ee
where $G=1+R_f$ is called {\em gain}, therefore as far as $0>R_f>\infty$ (thus retro-action is present) the device is amplifying.
\newline
Let us emphasize deep structural analogies with the Curie-Weiss response to a magnetic field $h$: First, we notice
that all these systems {\em saturate}.  Whatever the magnitude of the external field for the CW model, once all the spins become aligned,  increasing further $h$ will no longer produce a change in the system; analogously, once in the op-amp reached $V_{out} = V_{sat}$, larger values of $V_{in}$ will not result in further amplification. Also, notice that both the self-consistency and the transfer function are two input-output relations (the input being the external field in the former and the input-voltage in the latter, the output being the magnetization in the former and the output-voltage in the latter), and, once fixed $\beta=1$ for simplicity, expanding $\langle m \rangle = \tanh( J\langle m \rangle + h) \sim (1+J)h$, we can compare term by term the two expression as
\begin{eqnarray}
V_{out} &=&(1 + R_f) V_{in},\\
\langle m \rangle &=& (1 + J) h.
\end{eqnarray}
We see that $R_f$ plays as $J$, and, consistently, if $R_f$ is absent the retroaction is lost in the op-amp and the gain is no longer possible; analogously if $J = 0$, spins do not mutually interact and no feed-back is allowed to drive the phase transition.
\newline
Such a bridge is robust as operational amplifiers perform more than signal amplifying; for instance they can perform as latches, namely analog to digital converters. Latches can be achieved within the Curie-Weiss theory simply working in the low-noise limit as the sigmoidal function ($\langle m \rangle$ versus $h$, see Fig.$3$) of the self-consistency approaches a Heavyside, hence, while analogically varying the external field $h$, as soon as it crosses the value $h=0$ (say from negative values), the magnetization jumps discontinuously from $\langle m \rangle = -1$ to $\langle m \rangle = +1$ hence coding for a digitalization of the input.

\begin{figure}
\includegraphics[width=0.4\textwidth]{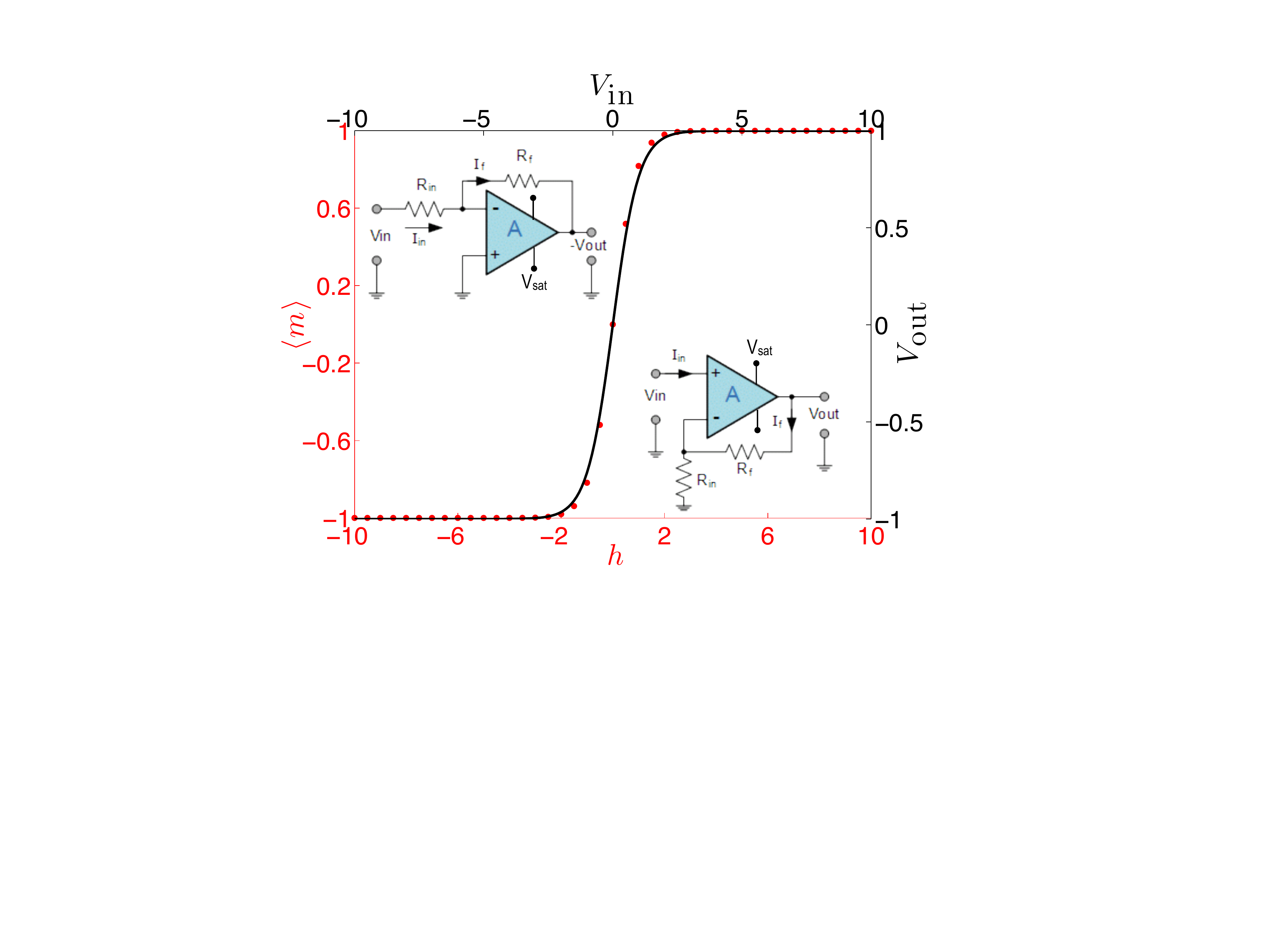}\label{fig:opamp}
 \caption{Average magnetization $\langle m \rangle$ versus the external field $h$ and response of a charging neuron (solid black line), compared with the transfer function of an operational amplifier (red bullets) \cite{21,SciRep}. In the inset we show a schematic representation of an operational amplifier (upper) and of an inverter (lower).}
\end{figure}

\subsection{The route from Curie-Weiss to Hopfield}
Actually, the Hamiltonian (\ref{CWH}) would encode for a rather poor model of neural network as it would account for only two stored patterns, corresponding to the two possible minima (that in turn would represent pathological network's behavior with all the neurons contemporarily completely firing of completely silenced), moreover, these ordered patterns, seen as information chains, have the lowest possible entropy and, for the Shannon-McMillan Theorem, in the large $N$ limit\footnote{The {\em thermodynamic limit} $N \to \infty$ is required for both mathematical convenience, e.g. it allows saddle-point/stationary-phase techniques, and in order to neglect observable fluctuations by a central limit theorem argument.} they will never be observed.

This criticism can be easily overcome thanks to the Mattis-gauge, namely a re-definition of the spins via $\sigma_i \to \xi_i^1 \sigma_i$, where $\xi_i^1= \pm 1$ are random entries extracted with equal probability and kept fixed in the network (in statistical mechanics these are called {\em quenched} variables to stress that they do not contribute to thermalization, a terminology reminiscent of metallurgy \cite{MPV}). Fixing $J \equiv 1$ for simplicity, the Mattis Hamiltonian reads as
\be
H_N^{Mattis}(\sigma|\xi) = -\frac{1}{2N} \sum_{i,j}^{N,N}\xi_i^1\xi_j^1\sigma_i \sigma_j - h\sum_i^N\xi_i^1 \sigma_i.
\ee
The Mattis magnetization is defined as $m_1 = \sum_i \xi_i^1 \sigma_i$. To inspect its lowest energy minima, we perform a comparison with the CW model: in terms of the (standard) magnetization, the Curie-Weiss model reads as $H_N^{CW} \sim -(N/2)m^2 - hm$ and, analogously we can write $H_N^{Mattis}(\sigma|\xi)$ in terms of Mattis magnetization as  $H_N^{Mattis} \sim -(N/2)m_1^2 - h m_1$. It is then evident that, in the low noise limit (namely where collective properties may emerge), as the minimum of free energy is achieved in the Curie-Weiss model for $\langle m \rangle \to \pm 1$, the same holds in the Mattis model for $\langle m_1 \rangle \to \pm 1$. However, this implies that now spins tend to align parallel (or antiparallel) to the vector $\xi^1$, hence if the latter is, say, $\xi^1 = (+1,-1,-1,-1,+1,+1)$ in a model with $N=6$, the equilibrium configurations of the network will be $\sigma=(+1,-1,-1,-1,+1,+1)$ and $\sigma=(-1,+1,+1,+1,-1,-1)$, the latter due to the gauge symmetry $\sigma_i \to -\sigma_i$ enjoyed by the Hamiltonian. Thus, the network relaxes autonomously to a state where some of its neurons are firing while others are quiescent, according to the {\em stored pattern} $\xi^1$. Note that, as the entries of the vectors $\xi$ are chosen randomly $\pm 1$ with equal probability, the retrieval of free energy minimum now corresponds to a spin configuration which is also the most entropic for the Shannon-McMillan argument, thus both the most likely and the most difficult to handle (as its information compression is no longer possible).

Two remarks are in order now. On the one side, according to the self-consistency equation (\ref{self}) and as shown in Fig.~$3$, $\langle m \rangle$ versus $h$ displays the typical graded/sigmoidal response of a charging neuron \cite{21}, and one would be tempted to call the spins $\sigma$ neurons. On the other side, it is definitely inconvenient to build a network via $N$ spins/neurons, which are further meant to be diverging (i.e. $N \to \infty$) in order to handle one stored pattern of information only. Along the theoretical physics route overcoming this limitation is quite natural (and provides the first derivation of the Hebbian prescription in this paper): If we want a network able to cope with $P$ patterns, the starting Hamiltonian should have simply the sum over these $P$ previously stored\footnote{The part of neural network's theory we are analyzing is meant for spontaneous retrieval of already stored information -grouped into patterns (pragmatically vectors)-. Clearly it is assumed that the network has already  overpass the learning stage.} patterns, namely
\be\label{hopfield}
H_N(\sigma|\xi) = -\frac{1}{2N}\sum_{i,j=1}^{N,N} \left(\sum_{\mu=1}^P \xi_i^{\mu}\xi_j^{\mu} \right) \sigma_i \sigma_j,
\ee
where we neglect the external field ($h=0$) for simplicity.
As we will see in the next section, this Hamiltonian constitutes indeed the Hopfield model, namely the harmonic oscillator of neural networks, whose coupling matrix is called {\em Hebb matrix} as encodes the Hebb prescription for neural organization \cite{3}.

\subsection{The route from Sherrington-Kirkpatrick to Hopfield}

Despite the extension to the case $P>1$ is formally straightforward, the investigation of the system as $P$ grows becomes by far more tricky. Indeed, neural networks belong to the so-called ``complex systems'' realm. We propose that complex behaviors can be distinguished by simple behaviors as for the latter the number of free-energy minima of the system \emph{does not scale} with the volume $N$, while for complex systems the number of free-energy minima \emph{does scale} with the volume according to a proper function of $N$. For instance, the Curie-Weiss/Mattis model has two minima only, whatever $N$ (even if $N \to \infty$), and it constitutes the paradigmatic example for a simple system.
As a counterpart, the prototype of complex system is the Sherrington-Kirkpatrick model (SK), originally introduced in condensed matter to describe the peculiar behaviors exhibited by real glasses \cite{4,MPV}. This model has an amount of minima that scales $\propto \exp(c N)$ with $c \neq f(N)$, and its  Hamiltonian reads as
\be
H_{N}^{SK}(\sigma|J) = \frac{1}{\sqrt{N}}\sum_{i<j}^{N,N}J_{ij}\sigma_i \sigma_j,
\ee
where, crucially, coupling are Gaussian distributed\footnote{Couplings in spin-glasses are drawn once for all at the beginning and do not evolve with system's thermalization, namely they are {\em quenched} variables too.} as $P(J_{ij})\equiv \mathcal{N}[0,1]$. This implies that links can be either positive (hence favoring parallel spin configuration) as well as negative (hence favoring anti-parallel spin configuration), thus, in the large $N$ limit, with large probability, spins will receive conflicting signals and we speak about ``frustrated networks''. Indeed {\em frustration}, the hallmark of complexity, is fundamental in order to split the phase space in several disconnected zones, i.e. in order to have several minima, or several stored patterns in neural network language. This mirrors a clear request also in electronics, namely the need for inverters (that once mixed with op-amps) result in flip-flops (crucial for information storage as we will see).

The mean-field statistical mechanics for the low-noise behavior of spin-glasses has been first described by Giorgio Parisi and it predicts a hierarchical organization of states and a relaxational dynamics spread over many timescales (for which we refer to specific textbooks \cite{MPV}). Here we just need to know that their natural order parameter is no longer the magnetization (as these systems do not magnetize), but the {\em overlap}  $q_{ab}$, as we are explaining. Spin glasses are balanced ensembles of ferromagnets and antiferromagnets (this can also be seen mathematically as $P(J)$ is symmetric around zero) and, as a result, $\langle m \rangle$ is always equal to zero, on the other hand, a comparison between two realizations of the system (pertaining to the same coupling set) is meaningful because at large temperatures it is expected to be zero, as everything is uncorrelated, but at low temperature their overlap is strictly non-zero as spins freeze in disordered but correlated states.
More precisely, given two ``replicas'' of the system, labeled as $a$ and $b$, their overlap $q_{ab}$ can be defined as the scalar product between the related spin configurations, namely as  $q_{ab} = (1/N)\sum_i^N \sigma_i^a \sigma_i^b$\footnote{Note that, while in the Curie-Weiss model, where $P(J)=\delta(J-1)$, the order parameter was the first momentum of $P(m)$, in the Sherrington-Kirkpatrick model, where $P(J)=\mathcal{N}[0,1]$, the variance of $P(m)$ (which is roughly $q_{ab}$) is the good order parameter.}, thus the mean-field spin glass has a completely random paramagnetic phase, with $\langle q \rangle \equiv 0$ and a ''glassy phase'' with $\langle q \rangle > 0$ split by a phase transition at $\beta_c =  T_c = 1$.

The Sherrington-Kirkpatrick model displays a large number of minima as expected for a cognitive system, yet it is not suitable to act as a cognitive system because its states are too ''disordered". We look for an Hamiltonian whose minima are not purely random like those in SK, as they must represent ordered stored patterns (hence like the CW ones), but the amount of these minima must be possibly extensive in the number of spins/neurons $N$ (as in the SK and at contrary with CW), hence we need to retain a ``ferromagnetic flavor''  within a ``glassy panorama'': we need {\em something in between}.

Remarkably, the Hopfield model defined by the Hamiltonian (\ref{hopfield}) lies exactly in between a Curie-Weiss model and a Sherrington-Kirkpatrick model. Let us see why: When $P=1$ the Hopfield model recovers the Mattis model, which is nothing but a gauge-transformed Curie-Weiss model. Conversely, when $P \to \infty$, $(1/\sqrt{N})\sum_{\mu}^P \xi_i^{\mu}\xi_j^{\mu}\to \mathcal{N}[0,1]$, by the standard central limit theorem, and the Hopfield model recovers the Sherrington-Kirkpatrick one. In between these two limits the system behaves as an associative network \cite{30}.
\newline
Such a crossover between CW (or Mattis) and SK models, requires for its investigation both the $P$ Mattis magnetization $\langle m_{\mu} \rangle$, $\mu = (1,...,P)$ (for quantifying retrieval of the whole stored patterns, that is the {\em vocabulary}), and the two-replica overlaps $\langle q_{ab} \rangle$ (to control the glassyness growth if the vocabulary gets enlarged), as well as a tunable parameter measuring the ratio between the stored patterns and the amount of available neurons, namely $\alpha=\lim_{N \to \infty}P/N$, also referred to as \emph{network capacity}.

\begin{figure}
 \includegraphics[width=0.385\textwidth]{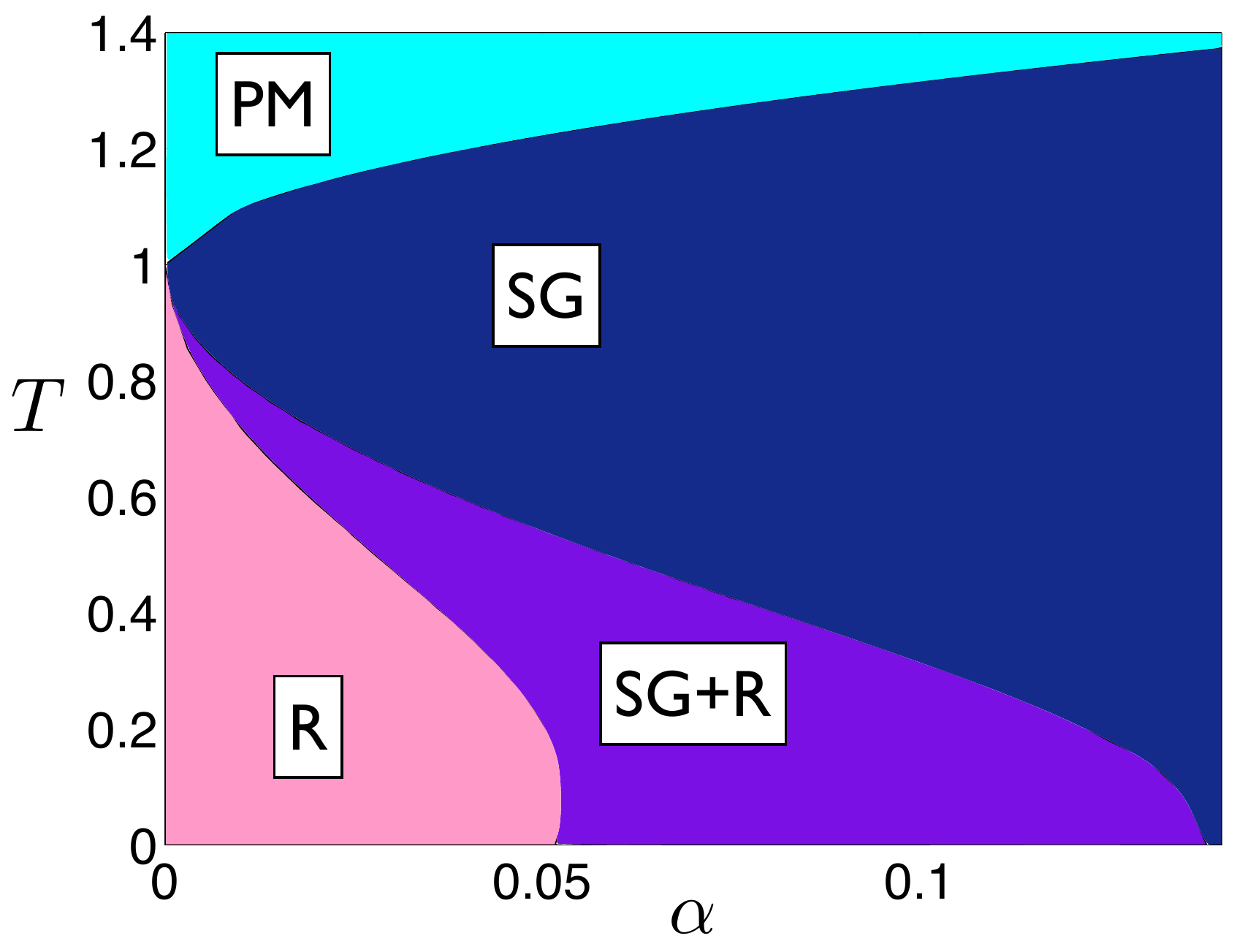}
 \caption{Phase diagram for the Hopfield model \cite{3}. According to the parameter setting, the system behaves as a paramagnet (PM), as a spin-glass (SG), or as an associative neural network able to perform information retrieval (R). The region labeled (SG+R) is a coexistence region where the system is glassy but still able to retrieve.}
\end{figure}

As far as $P$ scales sub-linearly with $N$, i.e. in the low storage regime defined by $\alpha=0$, the phase diagram is ruled by the noise level $\beta$ only: for $\beta<\beta_c$ the system is a paramagnet, with $\langle m_{\mu} \rangle =0$ and $\langle q_{ab} \rangle =0$, while for $\beta > \beta_c$ the system performs as an attractor network, with $\langle m_{\mu}\rangle \neq 0$ for a given $\mu$ (selected by the external field) and $\langle q_{ab}\rangle =0$. In this regime no dangerous glassy phase is lurking, yet the model is able to store only a tiny amount of patterns as the capacity is sub-linear with the network volume $N$.
\newline
Conversely, when $P$ scales linearly with $N$, i.e. in the high-storage regime defined by $\alpha >0$, the phase diagram lives in the $\alpha,\beta$ plane (see Fig.~$4$).
When $\alpha$ is small enough the system is expected to behave similarly to $\alpha=0$ hence as an associative network (with a particular Mattis magnetization positive but with also the two-replica overlap slightly positive as the glassy nature is intrinsic for $\alpha > 0$). For $\alpha$ large enough ($\alpha > \alpha_c (\beta), \alpha_c (\beta \rightarrow \infty) \sim 0.14$) however, the Hopfield model collapses on the Sherrington-Kirkpatrick model as expected, hence with the Mattis magnetizations brutally  reduced to zero and the two-replica overlap close to one. The transition to the spin-glass phase is often called ``blackout scenario'' in neural network community.
Making these predictions quantitative is a non-trivial task in statistical mechanics and, nowadays several techniques are available, among which we quote the replica-trick (originally used by the pioneers Amit-Gutfreund-Sompolinsky \cite{acab}), the martingale method (originally developed by Pastur, Sherbina and Tirozzi \cite{tirozzi}) and the cavity field technique (recently developed by Guerra and some of us in \cite{31}).

\begin{figure}
\includegraphics[width=0.4\textwidth]{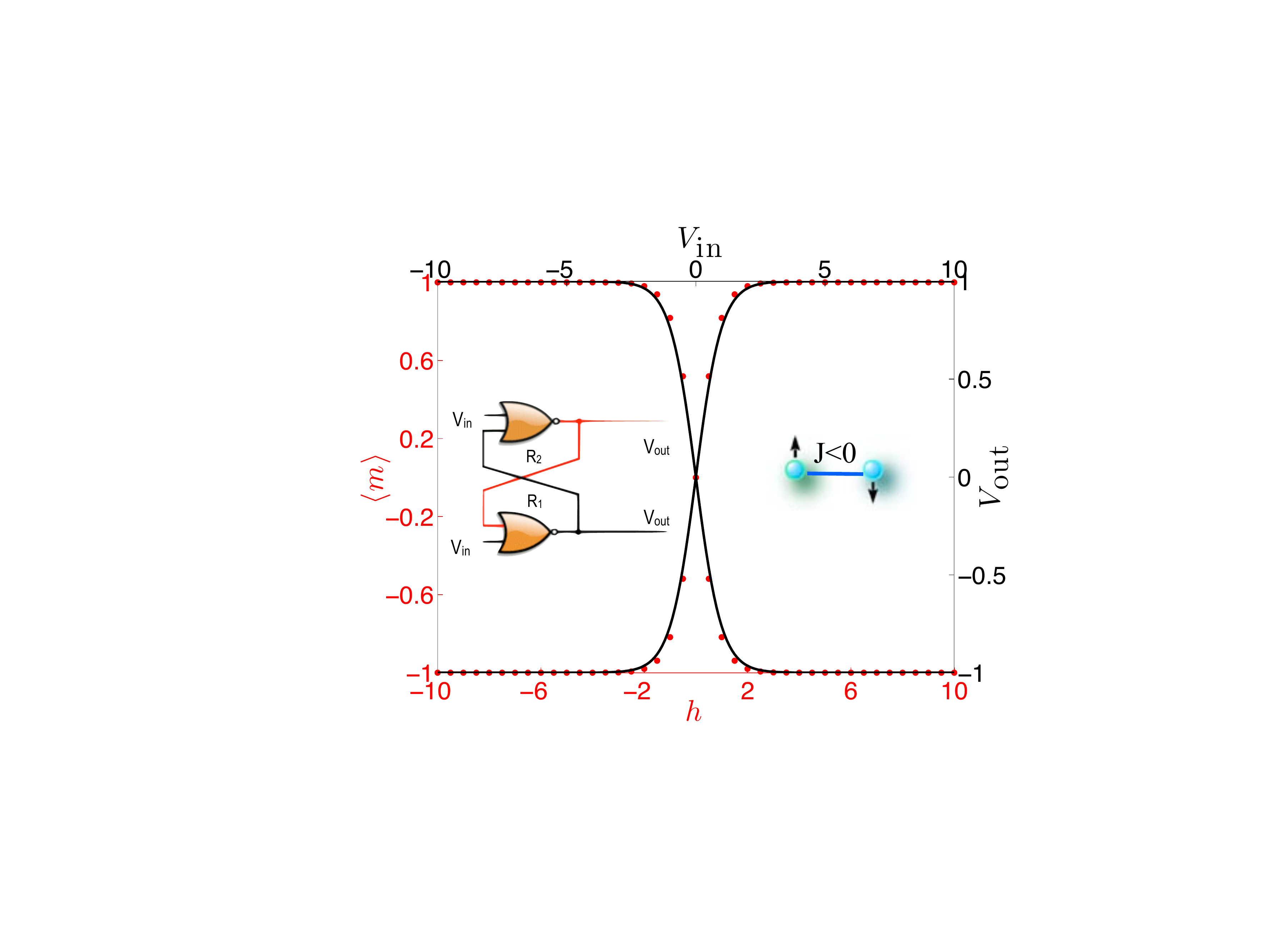}\label{fig:opamp}
 \caption{Average magnetizations $\langle m \rangle$ for each party of a bipartite antiferromagnetic model in the presence of the external field $h$ (solid black line), compared with the transfer function of  flip-flop (red bullets) \cite{21,SciRep}. In the left inset we show a schematic representation of a flip-flop while in the right inset we show a schematic representation of an antiferromagnetic coupling.}
\end{figure}

\subsection{Frustration and flip-flops}

As we saw, in order to store multiple patterns, the free energy landscape must be split into several valleys and this can be achieved only by introducing negative couplings (frustrating the network). This request mirrors the need for inverters (and -in turn- flip-flops) in synthetic information storage systems as we are going to sketch.  Because so far ferromagnetism has been already accounted (with its formal equivalence to the behavior of op-amps), now we turn to discuss antiferromagnetic behavior (which mathematically and conceptually mimics -in structure of matter- the behavior of flip-flops in electronics).

As a general scenario will then be the result of merging ferromagnetic and antiferromagnetic couplings in statistical mechanics and operational amplifiers and flip-flops in electronics, let us finally consider just a small subset of the whole Hopfield network, namely two groups of spins $A,B$ antiferromagnetically interacting (i.e. with a negative coupling, say  $J=-1$ instead of $J=+1$ of ferromagnetism)  -as shown in Fig.~$5$ (right inset)- and coupled with an external field $h$.
\newline
Not surprisingly, remembering eq.$(5)$ and that here $J=-1$, the self-consistencies coupled to the evolution of these two interacting spin groups read as
\begin{eqnarray}
\langle m_A \rangle &=& \tanh[\beta(-\langle m_B \rangle + h)],\\
\langle m_B \rangle &=& \tanh[\beta(-\langle m_A \rangle + h)].
\end{eqnarray}
The behavior of this system is shown still in Fig.$5$, where the transfer function of a flip-flop is pasted too for highlighting the structural equivalence.
\newline
An ensemble of spins that interact with both positive and negative couplings is a frustrated network whose free-energy decomposes in multiple valleys, namely a spin-glass. A particular spin-glass is the Hopfield model, where the coupling matrix has both positive and negative entries, but they obey the Hebb prescription. This system can be seen as an ensemble of ferromagnetic (positive coupling) and antiferromagnetic (negative coupling) interactions, thus, from an engineering perspective, as a linear combination of op-amps to amplify external signals and flip-flops to store the information they convey.

\subsection{Hopfield networks and Boltzmann machines}

The world of neural networks is nowadays very broad, with several models for several tasks. Hopfield networks are key models for retrieval, while Boltzmann machines are fundamental systems for learning \cite{BM}. However, learning and retrieval are not two independent operations, but, rather, two complementary aspects of cognition. Remarkably, the related models reflect this hidden link in fact, as we are going to show, Hopfield networks and Boltzmann machines obey two inseparable aspects of the same complex thermodynamics, hence it must be possible to recover the Hebb rule for learning also starting from (restricted) Boltzmann machines.
\newline
The latter, in their simplest development, are by-layered networks, with no links within any layer and only links among neurons of different layers (see Fig.~$6$). Links can be either positive or negative, hence also Boltzmann machines hide frustration and belongs to complex systems \cite{barra1}. If we use the symbol $\sigma_i$, $i \in (1,...,N)$, for neurons of one layer, $z_{\mu}$, $\mu \in (1,...,P)$ for neurons of the other layer, and $\xi_i^{\mu}$ to label the link between the spin/neuron (or node) $i$ and the spin/neuron (or node) $\mu$, we can write an Hamiltonian for the Boltzmann machine as
\be \label{BSG}
H_N(\sigma,z|\xi) = -\frac{1}{\sqrt{N}}\sum_{i=1}^N \sum_{\mu=1}^P \xi_i^{\mu} \sigma_i z_{\mu}.
\ee
The Hamiltonian (\ref{BSG}) represents, in the jargon of statistical mechanics, a bipartite spin-glass.
In order to study the related phase diagram we work out the statistical mechanics machinery and write the free energy of the Boltzmann machine as
\be
A(\beta)= \frac{1}{N}\log\sum_{\sigma}^{2^N}\sum_{z}^{2^P}\exp \left( \frac{\beta}{\sqrt{N}} \sum_{i=1}^N \sum_{\mu=1}^P \xi_i^{\mu}\sigma_i z_{\mu}\right).
\ee

\begin{figure}\begin{center}
 \includegraphics[width=0.4\textwidth]{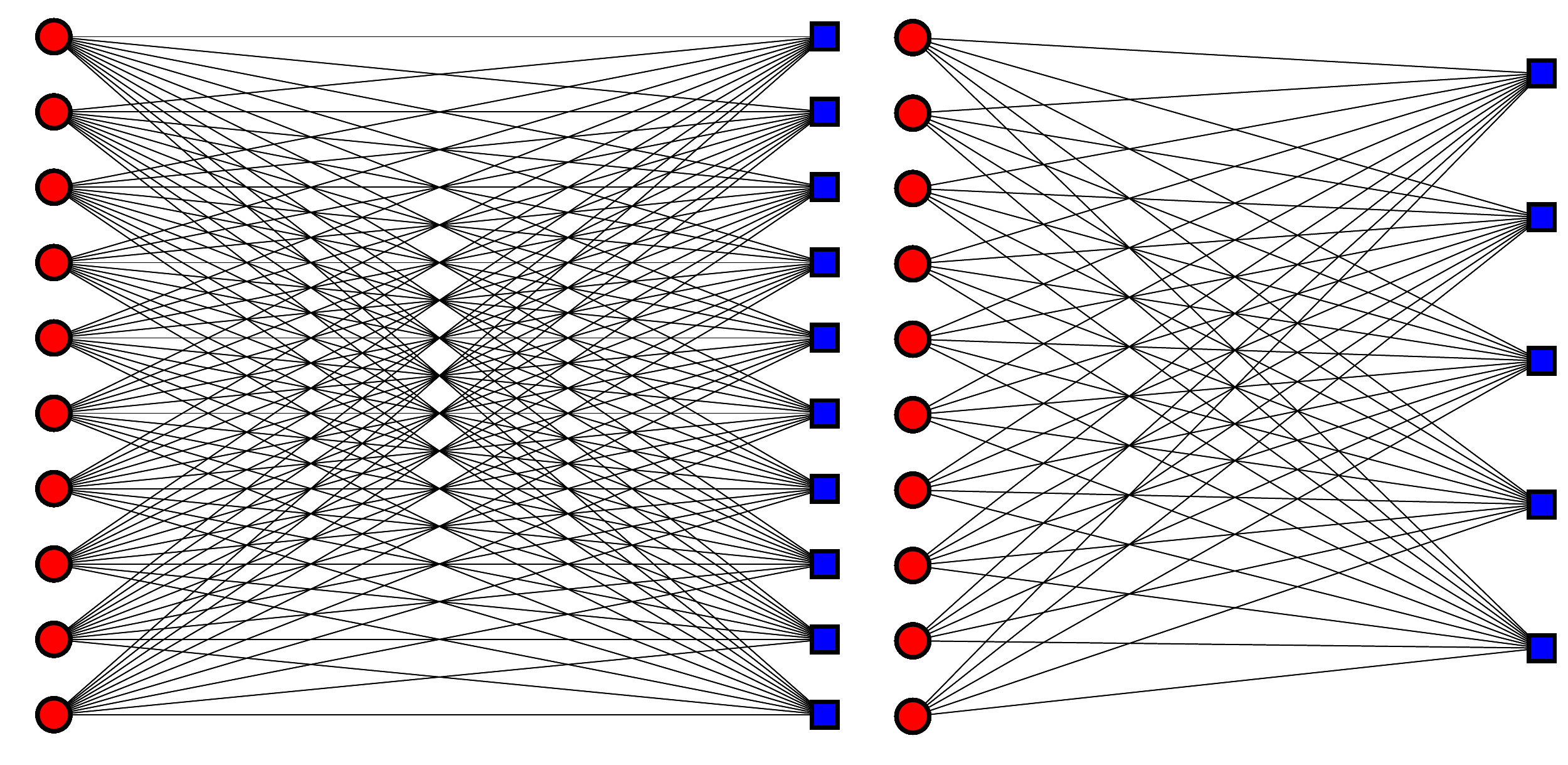}\end{center}
 \caption{Examples of Boltzmann machines, encoded by bipartite spin-glasses, with $\alpha=1$ (left panel) and $\alpha=0.5$ (right panel).}
\end{figure}

Remarkably, as there are no links within each party, from a statistical mechanics perspective, these networks are simple to deal with because the sums are factorized. In particular, we can carry out the sum over $z$ to get
\be \small
\nonumber
A(\beta)\sim \frac{1}{N}\log\sum_{\sigma}^{2^N}\exp \left [ \frac{\beta^2}{2N}\sum_{i,j}^{N,N} \left (\sum_{\mu=1}^P \xi_i^{\mu}\xi_j^{\mu} \right) \sigma_i \sigma_j \right ],
\ee
hence the leading contribution of the Boltzmann machine is nothing but the Hopfield network\footnote{Actually, with some additional mathematical efforts it can be shown that not only the leading term is an Hopfield network, but the whole machine behaves as  the Hopfield network.}. We remark that -as a sideline- we (re)-obtained the Hebb prescription for retrieval
also, from an artificial intelligence perspective, starting from a model meant for learning and, from a structure of matter perspective, from a spin-glass instead that from a ferromagnet.

\begin{figure} \label{fig:NMF}
 \includegraphics[width=0.4\textwidth]{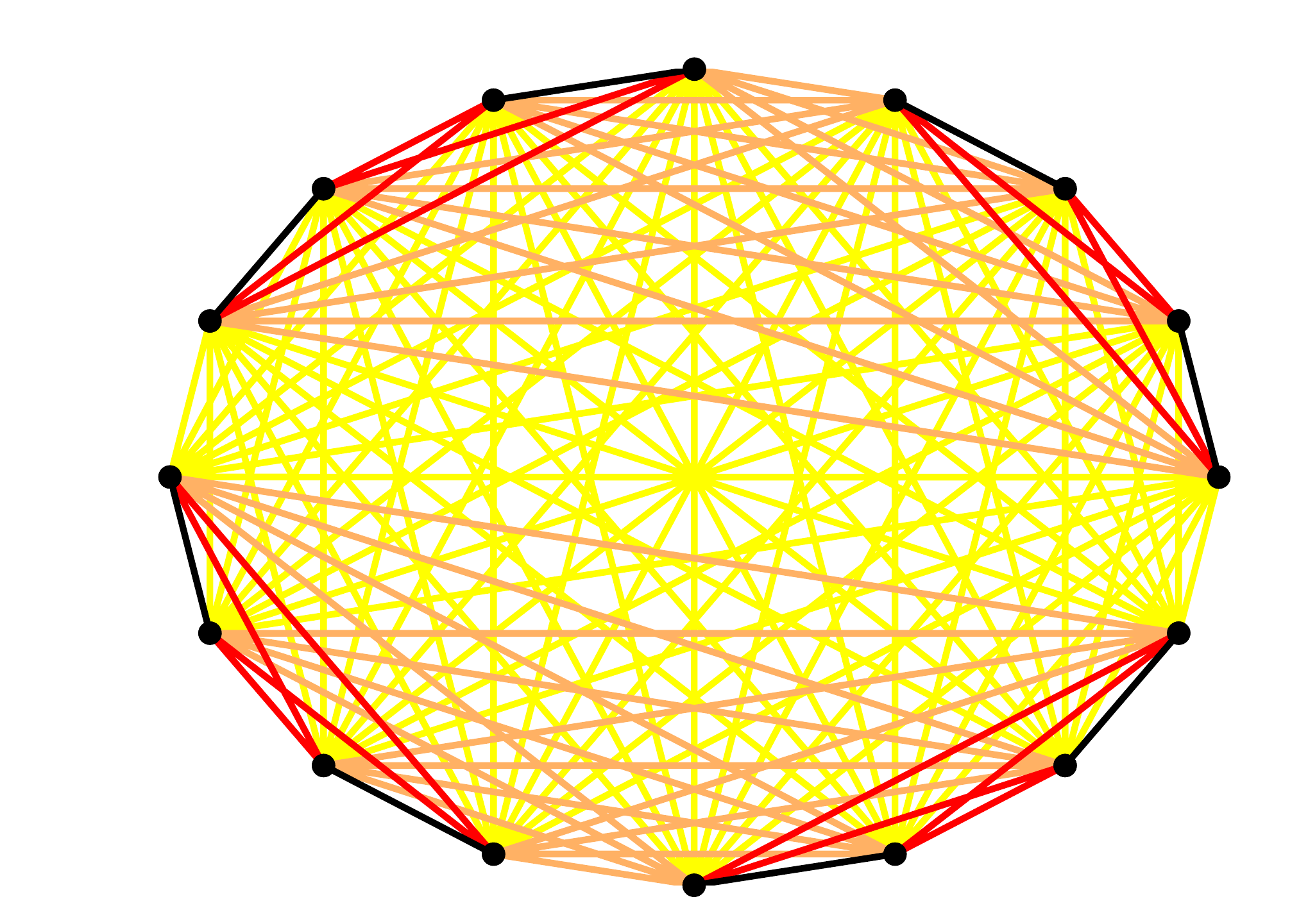}
 \caption{Example of a non-mean-field model where spins occupy nodes of a graph endowed with a metric distance. In this way, although each spin is connected with any other, the coupling strength decay as a power of distance, here represented with different shades: darker links correspond to stronger couplings. It can be useful a visual comparison with the mean-field topology shown in Fig.$1$ (right panel).}
\end{figure}

\section{Conclusions}

Proved by countless historically examples, research strongly benefits from interdisciplinary dialogues for at least a couple of reasons: one is clearly the unrestrainable formalization of soft-sciences by hard-ones as it is happening nowadays in systems biology. The other is that genuine ideas stemmed in a research branch can be suitably extended to cover aspects in a (usually closer or related) different research branches.
\newline
These notes are intended to contribute to the second route. Indeed, from the statistical mechanics perspective, the strand paved by engineers toward a theory for neural networks appears as the most {\em natural} way to proceed, or -in other words- the solely in perfect agreement with thermodynamic prescriptions.
\newline
Beyond tacking an historical perspective on such an evolution, framed within a modern theoretical physics scaffold, we gave here two alternative ways to (re)-obtain the celebrated Hebb rule, and we highlighted the thermodynamical equivalence between Hopfield networks and Boltzmann machines: remarkably, those models appear as a unique joint framework from theoretical physics, as learning and retrieval -their outcomes- are two inseparable aspects of the same phenomenon: spontaneous cognition.
\newline
Lastly, with the aim of driving engineer's curiosity toward statistical mechanics, we performed a one-to-one bridge between statistical mechanics outcomes and behaviors of electronic components: in particular we showed how the ferromagnetic scenario may play as a theory for signal amplification and how the antiferromagnetic scenario may play for information storage as happens in flip-flops. As ferromagnets and antiferromagnets synergically cooperate in glassy systems, and glassyness is intrinsic in neural networks theory in order to split the free energy into several minima thanks to  frustration, because we want to correspond each free energy minima with a stored pattern, op-amp and flip-flops must be crucial devices in practical realizations of neural networks: indeed they are.

We conclude with a remark about possible perspectives: we started this historical tour highlighting how, thanks to the mean-field paradigm, engineering (e.g. robotics, automation) and neurobiology have been tightly connected from a theoretical physics perspective. Now, however, as statistical mechanics is starting to access techniques to tackle complexity hidden even in non-mean-field networks (see e.g. Fig.~$7$, namely a hierarchical graph whose thermodynamics for the glassy scenario is almost complete \cite{giorgio}), we will probably witness another split in this smaller community of theoretical physicists working in spontaneous computational capability research: from one side continuing to refine techniques and models meant for artificial systems, well lying  in high-dimensional/mean-field topologies, and from the other beginning to develop ideas, models and techniques meant for biological systems only, strictly defined in finite-dimensional spaces or, even worst, on fractal supports.

\vspace{0.8cm}
\noindent
This work was supported by Gruppo Nazionale per la Fisica Matematica (GNFM), Istituto Nazionale d'Alta Matematica (INdAM).

\vfill

\bibliographystyle{apalike}
{\small
\bibliography{EngProc}}

\end{document}